\documentclass[conference]{IEEEtran}
\usepackage{cite}
\usepackage{amsmath,amssymb,amsfonts}
\usepackage{algorithmic}
\usepackage{graphicx}
\usepackage{textcomp}
\usepackage{xcolor}
\usepackage{subcaption}
\usepackage{hyperref}
\usepackage{balance}
\usepackage{booktabs}

\def\BibTeX{{\rm B\kern-.05em{\sc i\kern-.025em b}\kern-.08em
    T\kern-.1667em\lower.7ex\hbox{E}\kern-.125emX}}

\begin{document}
\title{High Performance Parallel I/O and In-Situ Analysis in the WRF Model with ADIOS2\\}

\makeatletter
\newcommand{\linebreakand}{
  \end{@IEEEauthorhalign}
  \hfill\mbox{}\par
  \mbox{}\hfill\begin{@IEEEauthorhalign}
}
\makeatother

\author{
  \IEEEauthorblockN{1\textsuperscript{st} Michael Laufer}
  \IEEEauthorblockA{
	\textit{Toga Networks, a Huawei Company}\\
    Tel Aviv, Israel \\
    michael.laufer@toganetworks.com}
  \and
  \IEEEauthorblockN{2\textsuperscript{nd} Erick Fredj}
  \IEEEauthorblockA{\textit{Computer Science Department}\\
 	\textit{The Jerusalem College of Technology}\\
 	Jerusalem, Israel \\
 	fredj@jct.ac.il\\
 	\textit{Toga Networks, a Huawei Company}\\
 	Tel Aviv, Israel \\
 	erick.fredj@toganetworks.com
 	}
}


\maketitle

\begin{abstract}
  As the computing power of large-scale HPC clusters approaches the Exascale, the gap between compute capabilities and storage systems is ever widening. In particular, the popular High Performance Computing (HPC) application, the Weather Research and Forecasting Model (WRF) is being currently being utilized for high resolution forecasting and research which generate very large datasets, especially when investigating transient weather phenomena. However, the I/O options currently available in WRF have been found to be a bottleneck at scale.

  In this work, we demonstrate the impact of integrating a next-generation parallel I/O framework - ADIOS2, as a new I/O backend option in WRF. First, we detail the implementation considerations, setbacks, and solutions that were encountered during the integration. Next we examine the results of I/O write times and compare them with results of currently available WRF I/O options. The resulting I/O times show over an order of magnitude speedup when using ADIOS2 compared to classic MPI-I/O based solutions. Additionally, the node-local burst buffer write capabilities as well as in-line lossless compression capabilities of ADIOS2 are showcased, further boosting performance. Finally, usage of the novel ADIOS2 in-situ analysis capabilities for weather forecasting is demonstrated using a WRF forecasting pipeline, showing a seamless end-to-end processing pipeline that occurs concurrently with the execution of the WRF model, leading to a dramatic improvement in total time to solution.
\end{abstract}


\section{Introduction}
The scientific applications for High Performance Computing (HPC) face a growing I/O challenge. Unfortunately, this increases demands on HPC storage subsystems, which leads to bottlenecks in simulation pipelines such as weather forecasting models. In this study we investigate the integration and application of a high-performance I/O and data management library - ADIOS2\cite{adios2}, with the now ubiquitous HPC application WRF (Weather Research and Forecasting Model)\cite{Skamarock08adescription}.

The paper is structured as follows: in section \ref{section:related} we will briefly discuss related works including WRF I/O studies, I/O libraries and related ADIOS2 HPC application I/O speedup achievements. Section \ref{section:Background} discusses the background of WRF as well as the ADIOS2 library used in this work. Section \ref{section:design} then details the implementation and challenges of integrating a new I/O library into WRF. Section \ref{section:Results} will then discuss the results of a set of I/O performance evaluations comparing ADIOS2 to other available WRF I/O methods, as well as an example in-situ analysis pipeline performance study, showcasing the significant wall time savings afforded by this work. Finally in section \ref{section:conclusions} we will present conclusions and next steps of this work.

\section{Related Work}
\label{section:related}
A number of previous studies have looked at the I/O scaling and bottlenecks in WRF. In the study conducted by Kyle\cite{akira} in partnership with NCAR, I/O time (file writes) was found to surpass compute time at scale on the Cheyenne and Yellowstone supercomputers running the Hurricane Maria 1km test case when process counts were raised to above 2000 compute cores.

Balle et al.\cite{balle2016improving} also shows that when using the PnetCDF I/O option, the write times increase as more nodes are brought online, and when the node count reaches about 500, the I/O time is measured at 50\% of the total run time. In that study the WRF Quilt Server functionality is used successfully to bring down I/O time, but at the cost of dedicated computational resources.

In Finkenrath et al.\cite{finkenrath}, testing of the default WRF I/O option - NetCDF, resulted in poor performance at scale. The Parallel NetCDF (PnetCDF) option was found to offer an order of magnitude speedup compared to its serial based cousin. Additionally, the researchers found that running WRF in hybrid MPI+OpenMP mode (\emph{dmpar+smpar}), greatly decreases the I/O time, due to the lower MPI communication costs between processes and associated file locking contention issues.

A similar study to this present work was carried out for the GRAPES mesoscale Numerical Weather Prediction application\cite{grapes}, where they integrated the first version of ADIOS as an I/O backend, and decreased the I/O time by an order of magnitude compared to the previous MPI-I/O based approach. In their work, they did not examine the in-situ pipelining capabilities of the ADIOS library, that are examined in this work.

Moreover, a previous work by Singhal and Sussman\cite{Singhal} integrated a version of ADIOS into WRF, with a focus on the development of a smart compression extension. Additionally, ADIOS was used to couple WRF with a second processing application. The program coupling was completed using traditional file open/read semantics and did not utilize the code coupling and transport capabilities that are now available in ADIOS2. Unfortunately, no upstream contribution was made to the WRF community, so this implementation could not be compared with the current work.

A number of studies have introduced I/O middleware libraries. MPI-I/O \cite{Corbett95overviewof}, NetCDF\cite{netcdf}, Parallel-NetCDF (PnetCDF) \cite{Li2003}, HDF5 \cite{hdf5} can boost I/O performance using parallel I/O involving a number of participating processes. These libraries primarily aim to optimize I/O for use with remote parallel file system (PFS). While these do indeed greatly improve I/O performance compared to serial methods, their functionality and flexibility are limited as they focus solely on file based I/O, in contrast to the data-management capabilities of ADIOS2. Moreover, the general approach of MPI-I/O based methods such as these is to write to a single file on the PFS, this means that they cannot directly take advantage of emerging high-speed node-local storage (burst buffers), which are distributed on compute nodes\cite{Oral2019}. In our work we showcase the next generation I/O library, along with its multitude of features.

\section{Background}
\label{section:Background}
The following section details the background of both the WRF model and the ADIOS2 library.

\subsection{Weather Research Forecasting Model}
\subsubsection{WRF Background}
WRF is a state of the art mesoscale Numerical Weather Prediction system (NWP) intended both for forecasting and atmospheric research. It is an Open Source project, officially supported by the National Center for Atmospheric Research (NCAR), has become a true community model by its long-term development through the interests and contributions of a worldwide user base. The software framework of WRF has facilitated such extensions and supports efficient, massively-parallel computation across a broad range of computing platforms. The governing equation set of the WRF model is based on the compressible, non-hydrostatic atmospheric motion with multiple physics processes such cloud and precipitation, boundary layer turbulence, land ocean air interaction, radiative transfer in the atmosphere, and energy transfer at the surface. The finite difference method is used to discretize the governing equations of the WRF model. These discretized equations are integrated in time to obtain time-dependent atmospheric motion and physical states. Owing to the multiple physical processes that determine the atmospheric motion field, the number of prognostic variables of the WRF model is quite large compared to a simple Computational Fluid Dynamics (CFD) model that consists of the Navier-Stokes equation and the mass continuity equation. The large number of prognostic variables in the three dimensions is a severe computational as well as storage solution constraint, which requires high performance resources. WRF produces atmospheric simulations. The process has two phases, with the first to configure the model domain(s), ingest the input data, and prepare the initial conditions, and the second to run the forecast model and output solution and checkpoint files. The forecast model components operate within WRF’s software framework, which handles I/O and parallel-computing communications. WRF is written primarily in Fortran, can be built with a number of compilers, and runs predominately on platforms with UNIX-like operating systems, from laptops to supercomputers.

\subsubsection{WRF I/O Backends}
WRF’s well-defined I/O API provides several different implementations of its I/O layer, the ones relevant for the present work:

\begin{itemize}
\item  Serial NetCDF\cite{netcdf} (\emph{io\_form=2}): The default I/O option in the WRF model. When this I/O option is selected, all data is funneled through the first MPI rank, where this rank alone writes out a NetCDF4 based file using the NetCDF library (HDF5 based). While the root rank is writing to disk, all other ranks wait until the write has fully concluded before continuing computation. This method performs well at low process counts but at higher counts, the write time can quickly dominate the computation time. One of the main advantages of this method is the ability to use lossless compression that is integrated in NetCDF through the HDF5 library. This results in significantly smaller file sizes, achieving compression ratios close to 4 in the cases examined in this work. Still, due to the massive communication overhead, and single write thread, this option achieves poor I/O performance at scale.
\item Split NetCDF (\emph{io\_form=102}): This option also uses the NetCDF library for I/O but instead of sending all data to the first MPI rank, each rank writes its own distinct file. As will be seen later in this work, this method is able to achieve very high throughput at moderate MPI rank counts due to the absence of communication overhead, but this file-per-process method (\emph{N} processes to \emph{N} files, i.e \emph{N-N} I/O approach) does not scale to high process counts due to the immense pressure on the underlying file system and metadata servers. Additionally, as this output method generates multiple distinct files, the post processing is not trivial, especially when the rank of readers does not match the amount of files. To counter this, a community provided routine can stitch together the output files back into a single file, but this also incurs a non-negligible time penalty, as well as additional complexity in post processing pipelines.
\item Parallel NetCDF\cite{Li2003} (\emph{io\_form=11}): WRF's primary parallel I/O option that utilizes MPI-I/O. When this method is employed, all MPI ranks cooperate to write a single output file in parallel (\emph{N-1} I/O approach). As opposed to NetCDF4 based methods, this method does not currently allow for data compression. Even without compression capabilities, this option has been shown to offer an order of magnitude increase in write bandwidth compared to the serial NetCDF method at scale, due to the coordinated MPI-IO two-phase method\cite{finkenrath}. As this is the primary parallel I/O option that allows for operation without requiring additional overhead from stitching multiple files together, it is treated as the reference benchmark method (baseline), when comparing against the ADIOS2 method introduced in this work.
\item Quilt Servers: The quilt server technique uses dedicated I/O processes ("servers") that deal exclusively with I/O, enabling the compute processes to continue with their work without waiting for data to be written to disk before proceeding. Data from multiple compute ranks are merged ("quilted") together by a dedicated I/O rank by means of MPI communication calls and kept in system memory until they are written to PFS. This option is not investigated in this work, but should be investigated in future works.
\end{itemize}

\subsection{ADIOS2}
ADIOS2, is the second generation of the Adaptable Input Output System\cite{adios}, first introduced by Lofstead et al. The ``adaptability'' is attributed to the ability for the ADIOS libraries to use different I/O techniques, file formats, and even transports that are configured at run time using an XML configuration file, while remaining optimized for operation at scales from laptops to supercomputers. ADIOS2, while C++ based, provides bindings for additional languages such as C, Fortran, Python, and Matlab.

Importantly, ADIOS2 is more than just a file based I/O library such as Parallel NetCDF\cite{Li2003}, and HDF5\cite{hdf5}, as it enables in-situ analysis, Wide Area Network transport capabilities, as well as in-line data operations (e.g lossy/lossless compression, data transformations).

Still, high performance file based I/O is a primary goal of ADIOS2. Using its own file format, ADIOS2 designates a variable number, \emph{M}, of MPI ranks as aggregators that each write out their own sub-file, while collecting data in a streaming fashion from their designated MPI ranks (\emph{N-M} I/O approach). Data is transferred from neighboring rank to neighboring rank while the aggregator continually writes the data that it receives. As each aggregator writes its own distinct file, there is no chance for file locking issues that are prevalent with MPI-I/O (\emph{N-1} I/O approach) based approaches used in PnetCDF and HDF5. ADIOS2 then uses a smart metadata algorithm to keep track of where the data buffers are located within the sub-files, that is then used to reconstitute the data for reading.

Moreover, the number of aggregators and their placement is also tunable at runtime, and is the primary tuning knob for file based I/O in ADIOS2. This adaptable N-M writing approach, streaming data, and the absence of global data sharing between MPI ranks like in MPI-I/O, leads to orders of magnitude improvements to write bandwidth \cite{adios2}. 

Additionally, ADIOS2 offers node-local burst buffer functionality that writes the sub-files for each process on its own node-local storage device (burst buffer), and then uses a separate thread to drain the burst buffer data back to the Parallel File System while the host application can reuse the application data buffers and continue computation.

ADIOS2 is under continual development and has already been successfully integrated into a number of key HPC applications:
\begin{itemize}
	\item OpenFOAM\cite{openfoam}
	\item LAMMPS\cite{lammps}
	\item XGC\cite{XGC}
	\item E3SM\cite{E3SM}
	\item Trilinos\cite{trilinos}
	\item PETSc\cite{petsc}
\end{itemize}

ADIOS2 offers a number of different engines that can be selected at runtime. Aside from the file-based engines like \emph{BP4}, the primary engine that will be utilized in this work is Sustainable Staging Transport engine (SST). SST allows direct connection of data producers with consumers using the same ADIOS2 write/read API as the file based engines. This allows for data to fully bypass the filesystem, and reach the consumer (reader) using a high speed RDMA network interconnect. SST supports a variable number of producers and consumers, (they need not match in quantity), and will buffer data in the producer's memory until the consumer is ready to receive the data. This allows for seamless in-situ processing, where the computation can continue while analysis is being performed.

Lastly, ADIOS2 allows for in-line data operations on the data stream. One of the primary operations used is in-line data compression. ADIOS2 allows for both lossy and lossless compression using a number of compressors and codecs. This work utilizes the lossless compression capabilities that ADIOS2 offers through the Blosc\cite{blosc} meta-compressor.

\section{Design and Implementation}
\label{section:design}
The ADIOS2 I/O backend is implemented similarly as the other \emph{external} I/O options within WRF with some minor differences. The implementation is based on the distributed memory PnetCDF option, with the PnetCDF \emph{ncmpi\_put\_var\_type} calls replaced ADIOS2 \emph{adios2\_put} calls. 

One of the main differences between a NetCDF based approach and the ADIOS2 approach is how each takes into account the time dimension. NetCDF treats time as a distinct dimension in the global data arrays (i.e a fourth dimension when writing 3D data). ADIOS2, on the other hand treats the time dimensions differently than the physical dimensions. ADIOS2 data is organized by distinct time steps (\emph{step-based} approach). To accomplish this, ADIOS2 must know the start and end of data production/ingestion time steps. Therefore, one small change within the main WRF I/O logic loop was added to pass the start-time-step and end-time-step tags to the ADIOS2 I/O backend.

Additionally, ADIOS2 compression is generally defined at run time using an XML configuration file, and must be specified for each output variable. As the specific variables in WRF can change based on the physics suite selected, coupled with the overwhelmingly large possible number of output variables (sometimes over 200), adding these definitions manually into the XML configuration file is not feasible. Therefore, the compression settings are integrated within WRF through an additional option in the \emph{namelist.input} file.

Lastly, in order to accelerate adoption by the WRF community, who predominantly use NetCDF file formats for processing, a Python based stand-alone converter program was written to convert the WRF ADIOS2 format back into a NetCDF format. This allows for backwards compatibility with legacy post processing pipelines. Conversion time for history output files of the CONUS 2.5km model were found to be below 10 seconds using a single execution thread on a small workstation machine. 

\section{Results}
\label{section:Results}
In this section we evaluate the new ADIOS2 I/O backend within WRF, and compare it to other I/O options available in the model. Additionally, the ADIOS2 burst buffer write capabilities are investigated, as well as the effects of in-line compression on write times and output data sizes. Lastly, an ADIOS2 based, in-situ processing pipeline is demonstrated and end-to-end run times are compared to the traditional post processing method.

For the testing in this work a compute cluster was used consisting of 8 compute nodes, each equipped with two 18 core Intel Xeon Gold 6240 CPUs (288 total cores), and 384 GB DDR4 memory rated at 2933 MHz. For interconnect, a Mellanox ConnectX-6 using one port of 100GBe was used in each compute node.

In regards to storage, a dedicated storage node was outfitted with a BeeGFS parallel file system\cite{beegfs} striped over eight disks, and connected to the compute nodes using a Mellanox ConnectX-5 NIC. Lastly, each compute node was also equipped with an Intel DC P4510 1TB NVMe SSD drive with rated speeds of 2850 MB/s sequential read and 1100 MB/s sequential write.

For WRF testing, v4.2 was used, and configured for distributed memory mode (i.e \emph{dmpar}). In order to support the different WRF I/O options, WRF was compiled against NetCDF v4.8.1, PnetCDF v1.12.1 and ADIOS2 (Github, master branch\cite{adios2github}) using GCC v10.2.
A WRF test case were selected for I/O testing, the classic continental US, i.e "CONUS" benchmark at 2.5km XY spatial resolution. The CONUS 2.5km case, along with its less intensive CONUS 10km case are considered official benchmarks for WRF and are available on the WRF web site \url{https://www2.mmm.ucar.edu/wrf/WG2/bench/}. At the time of development, these official benchmarks were only compatible with WRF v3 and not WRF v4.2, used here. As such, the \emph{New} CONUS 2.5km benchmark was developed for WRF v4+ in a study by Kyle \cite{akira}, and was adopted for this work.

The WRF test case was modified, with the WRF history file frequency increased to one file every 30 simulation time minutes to represent a data analysis time scale that is relevant for time resolving transient weather analysis.

Test runs with up to 8 compute nodes (288 MPI ranks) were performed 5 times for each test configuration, and average I/O times were computed for each test.

\subsection{ADIOS2 File Write Performance}
Three I/O configurations were evaluated and run with the CONUS 2.5km model, while using the BeeGFS file system as the storage target. First the baseline parallel I/O option - PnetCDF, followed by the Split NetCDF method, and finally the new ADIOS2 method. The performance of the serial NetCDF I/O option was not tested, as it is known to not perform adequately at high process counts\cite{finkenrath}. Average WRF history file write results for the CONUS 2.5km benchmark case can be seen in Fig. \ref{fig:adios2_comparison} for a number of node counts, demonstrating the scaling performance of each I/O option.

\begin{figure*}[htbp]
	\centerline{\includegraphics[]{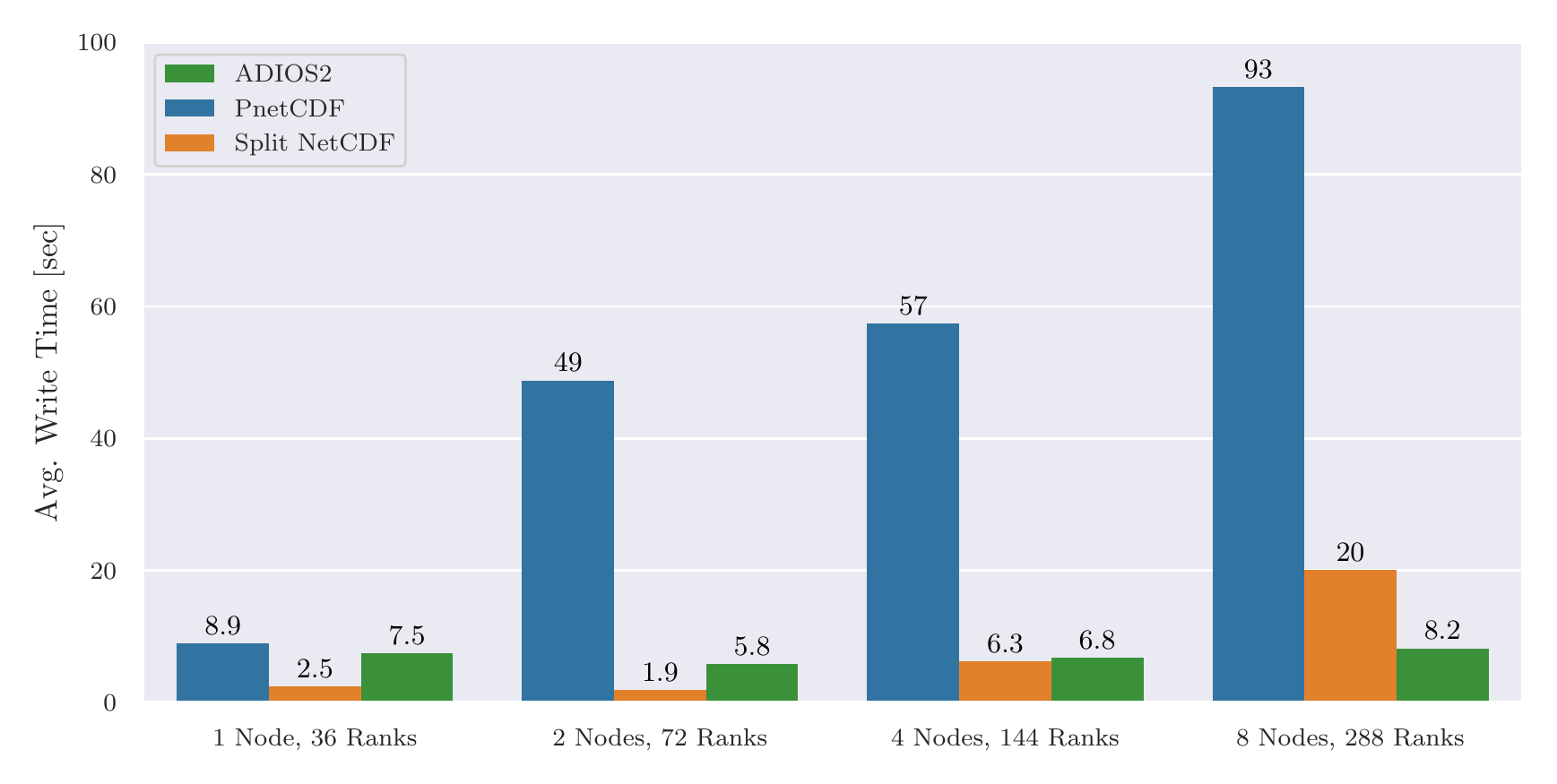}}
	\caption{Comparison of average history file write times of ADIOS2 compared to legacy parallel I/O options in WRF for different node/rank counts for the CONUS 2.5km model. ADIOS2 shows over an order of magnitude improvement over the baseline PnetCDF using 8 nodes, 288 ranks.}
	\label{fig:adios2_comparison}
\end{figure*}

The PnetCDF write times can be seen rising as nodes are added, this is due to the additional global inter-node communication required for the two-phase MPI-I/O based method employed by the PnetCDF library implementation. The Split NetCDF method shows impressive results at low node counts, but between 4 and 8 nodes, the write time rises by a factor of 3, exhibiting poor scaling performance, which may be caused by file system contention.

By contrast, the ADIOS2 method shows the most consistent results across the process count range, as write times remain stable as nodes are added. ADIOS2 bests the baseline PnetCDF results by an order of magnitude, while also achieving a relative write time speedup of over 2X compared to the Split NetCDF method at 8 compute nodes.

\subsection{ADIOS2 Burst Buffer}
To investigate the performance of the ADIOS2 burst buffer feature, the ADIOS2 XML configuration file was adjusted to target the node-local NVMe SSD on each compute node. In this configuration, the ADIOS2 aggregators on each node write their data locally and a background thread can be used to drain the burst buffer contents back to the PFS while the computation continues to progress. For this set of testing the drain feature was disabled.

A comparison of the scaling performance of the ADIOS2 burst buffer compared against the normal PFS write can be seen in Fig. \ref{fig:adios2_bb_comparison}. Compared to the normal PFS write configuration, the burst buffer results exhibit similar times to the PFS write at low node counts, but show a dramatic decrease in average write time as more nodes are added. This is due to the supplemental write bandwidth of the node local NVMe SSDs on each additional node. Additionally the speedup of the ADIOS2 burst buffer configuration compared to the burst buffer results of a single node is plotted in Fig. \ref{fig:adios2_bb_scaling}. The results show ideal write time scaling up to 4 nodes with only a small deviation from ideal at 8 nodes. This is in stark contrast to the MPI-I/O based results of PnetCDF showing an inverse speedup trend as more nodes are added.

The ADIOS2 burst buffer functionality greatly accelerates I/O write performance, and in this case shows a \emph{two} order of magnitude speedup compared to the benchmark WRF parallel I/O option, PnetCDF, on the system used in this work. As PnetCDF writes to a single file, it cannot directly utilize node-local storage to accelerate I/O.

\begin{figure}[htbp]
	\centerline{\includegraphics[]{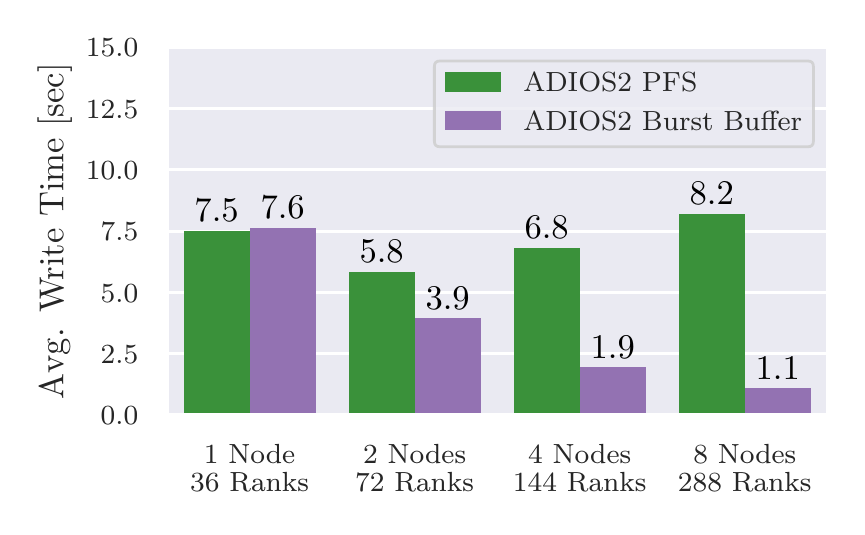}}
	\caption{Comparison of average history file write times of ADIOS2, when using the node-local burst buffer feature compared to writing to the PFS, for the CONUS 2.5km model.}
	\label{fig:adios2_bb_comparison}
\end{figure}

\begin{figure}[htbp]
	\centerline{\includegraphics[]{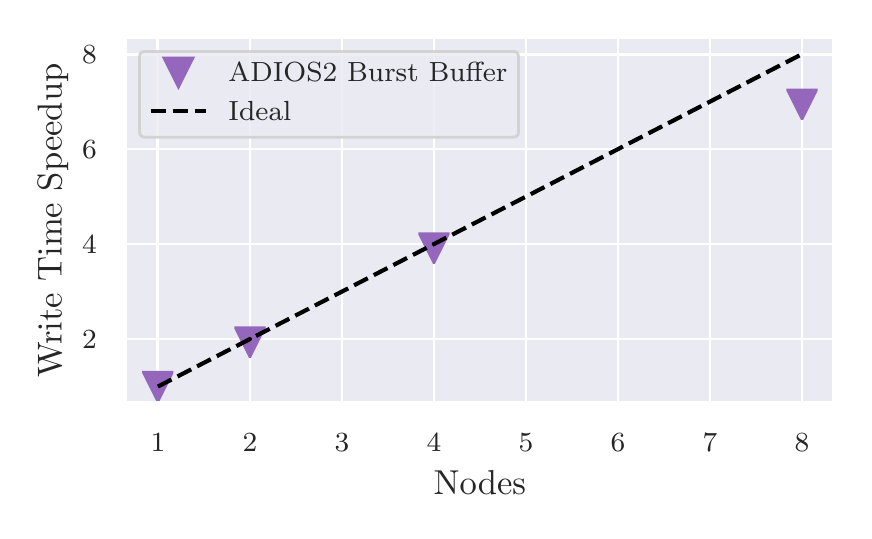}}
	\caption{Average history write time speedup when using the ADIOS2 burst buffer feature as a function of the amount of compute nodes. Scaling is found to closely achieve ideal values.}
	\label{fig:adios2_bb_scaling}
\end{figure}

\subsection{ADIOS2 Number of Aggregator}
The primary tuning knob for ADIOS2 file based I/O is the number aggregators, which dictates the number of sub-files that are written independently to the file system. In current versions of ADIOS2, the default behavior is to use a single aggregator per node, whereby one MPI rank acts as the aggregator for the rest of the ranks on the node. Based on capabilities of the underlying filesystem and problem size, an alternative number of aggregators/sub-files may be ideal. ADIOS2 allows the number of aggregators to be chosen at run time, in order to optimize the output speeds. In this next set of tests, the effect of the number of aggregators is investigated.

Fig. \ref{fig:adios2_aggregator} displays the effect of the number of aggregators per node on the average history write time for the CONUS 2.5km case for a single compute node, as well as 8 compute nodes. We can observe that for a single compute node, a large number of aggregators leads to substantially lower write times, while for the 8 compute node scenario, the optimal configuration is a single aggregator per node. This is consistent with the results seen when comparing the serial NetCDF and Split NetCDF results, where at high node counts, the Split NetCDF results showed a deterioration with regards to the write time. In contrast, ADIOS2 gives the user the freedom to choose the optimal amount of aggregators (sub-files) for the case at hand. In general the optimal number of aggregators is a function of the underlying file system, problem size, and amount of MPI ranks. The implementation of the ADIOS2 backend in WRF allows for the number of aggregators to be defined at runtime in the\emph{namelist.input} file. For all subsequent tests using 8 compute nodes, the number of aggregators/sub-files was defined as 1 per node.

\begin{figure}[htbp]
	\centerline{\includegraphics[]{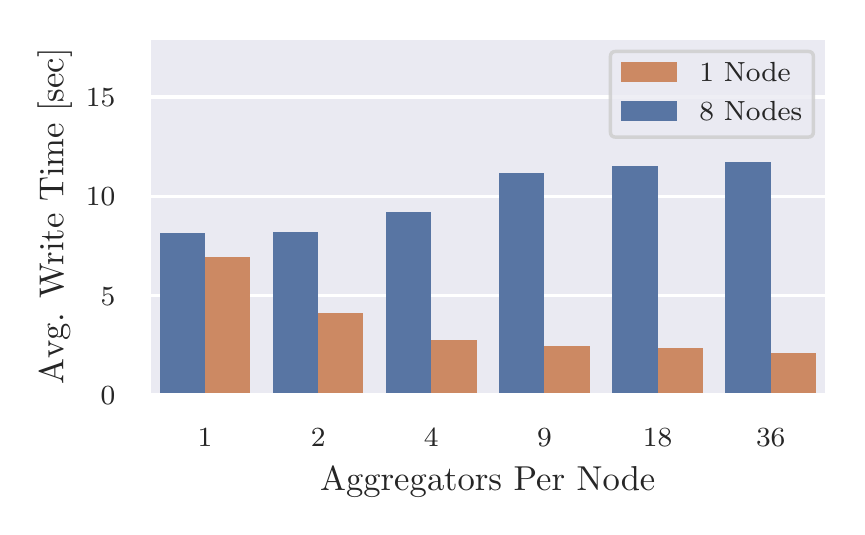}}
	\caption{Comparison of the average history write time for the CONUS 2.5km model, while varying the number of ADIOS2 aggregators per node. The optimal number of aggregators per node is found to be case dependent. }
	\label{fig:adios2_aggregator}
\end{figure}

\subsection{ADIOS2 in-line Compression}
Using the Operator abstraction within ADIOS2, data can be manipulated in-line. One of the primary operator classes is the use of in-line compression of the data stream. This compression can be achieved using a number of available lossy and lossless compression backends and codecs. In this work, the Blosc\cite{blosc}, ``meta-compressor'' was selected as it is lossless and supports a number of state-of-art compression codecs.
The following Blosc compression codecs were tested:
\begin{itemize}
	\item BloscLZ\cite{blosc}
	\item LZ4\cite{lz4}
	\item Zlib\cite{zlib}
	\item Zstd\cite{zstd}
\end{itemize}
Fig. \ref{fig:adios2_compression} shows a write performance scaling comparison of the average history file write times of compressed ADIOS2 data for a number of Blosc compression codecs, compared to uncompressed data when writing to the PFS. The test results show that an approximate speedup of 50\% is achieved when using ADIOS2 compression across the range of node/rank count. In particular, the Zstd\cite{zstd} codec takes the performance crown, as it achieves the lowest average history file write time in 3 out of 4 tests on this system.

\begin{figure*}[htbp]
	\centerline{\includegraphics[]{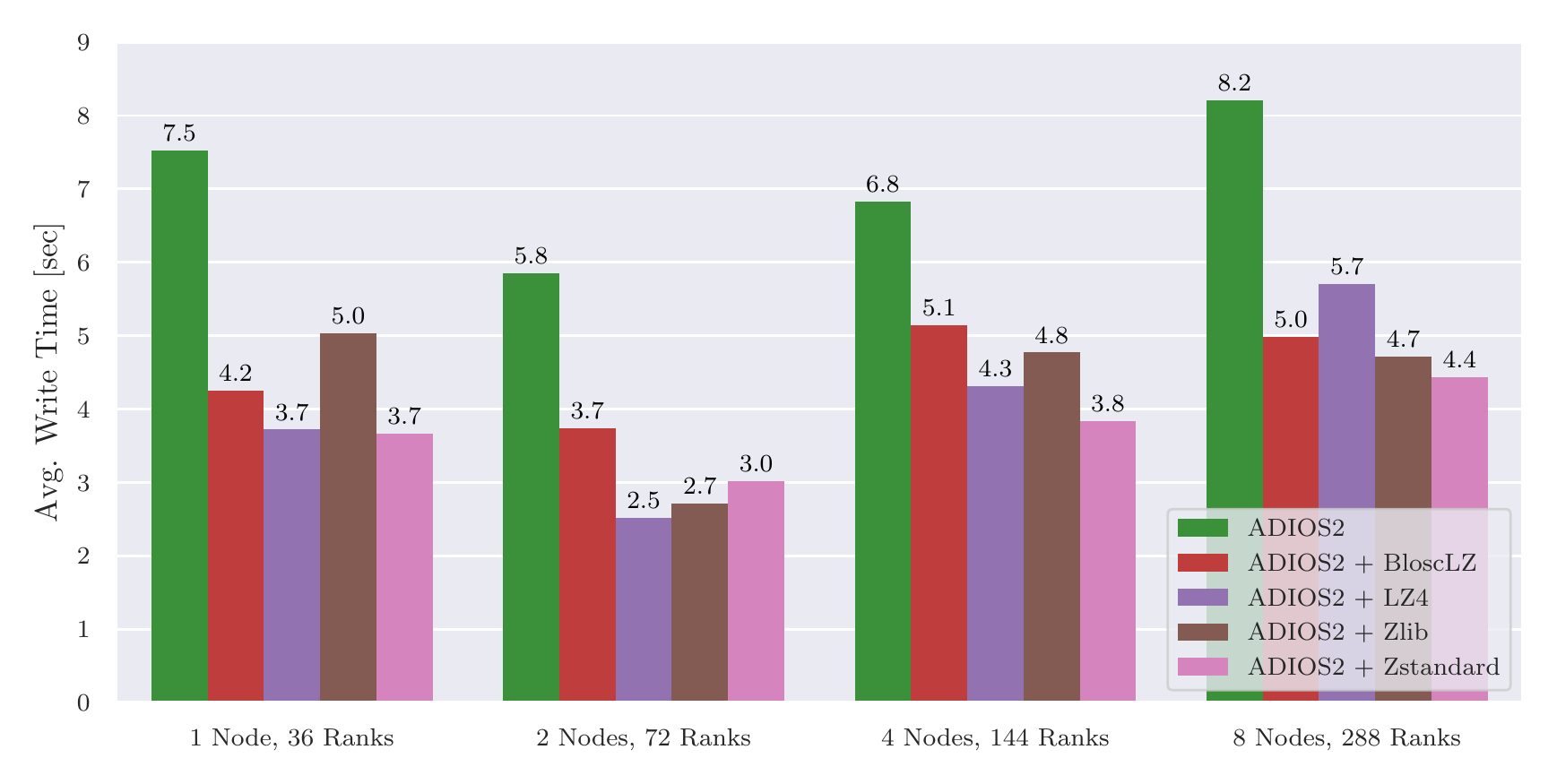}}
	\caption{Comparison of average history file write times of ADIOS2 uncompressed and ADIOS2 compressed data using different Blosc compressor codecs. The compressed ADIOS2 configurations enable close to a 50\% reduction in average write time compared to the raw ADIOS2 data configuration.}
	\label{fig:adios2_compression}
\end{figure*}

The output data size of a single CONUS 2.5km history file was recorded for the uncompressed ADIOS2, along with the size for each of the tested Blosc compression codecs. Additionally, the output sizes of the alternative WRF I/O options are shown. The  NetCDF4 based (Zlib\cite{zlib} compression), serial NetCDF I/O method file sizes were recorded along with the uncompressed NetCDF3 based PnetCDF output file size. Output data file size results can be seen in Fig. \ref{fig:adios2_compression_size}.

Clearly, the compression methods within both ADIOS2 (Blosc) and those utilized within the NetCDF4 WRF I/O implementation achieve impressive compression ratios of about 4, which greatly reduce the storage volume required. Moreover, aside from Zlib, the Zstd file size is smallest between the Blosc codecs, while also achieving maximal throughput. In test results not shown here, the LZ4 codec showed the most consistent compression throughput across a range of architectures, including \emph{AArch64}. The LZ4 compression codec was selected as the default codec in the WRF implementation, but can be configured at runtime in the \emph{namelist.input} file. 

\begin{figure}[htbp]
	\centerline{\includegraphics[]{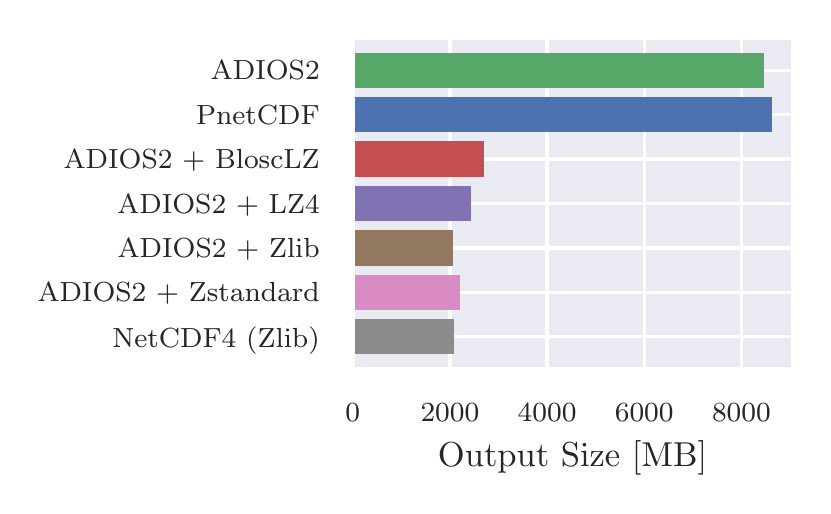}}
	\caption{Comparison of the compressed vs uncompressed output data size using different ADIOS2 compression codecs. }
	\label{fig:adios2_compression_size}
\end{figure}

\subsection{Optimal ADIOS2 File Write Configuration for WRF}
A separate test run using the optimal configuration discovered for ADIOS2 file writes on the testbed was carried out using compute 8 nodes. The node-local NVMe SSDs were chosen as the ADIOS2 storage target, and the Blosc compressor with the Zstd\cite{zstd} codec was selected, while using the an aggregator count of 1 per node. Results showing the effects of each subsequent optimization are shown in the Table \ref{tab:progression}, starting with the PnetCDF configuration writing to PFS.

\begin{table}[h!]
	\caption{Progression of Optimizations.}
	\centering
	\begin{tabular}{|l|c|c|}
	\hline
	Configuration & Write Time [s] & Speedup\\
	\hline
	PnetCDF             & 93                 & 1X      \\
	ADIOS2              & 8.2                & 11X     \\
	ADIOS2+BB           & 1.1                & 84X     \\
	ADIOS2+BB+Zstd      & 0.52               & 179X    \\ \bottomrule
	\hline
	\end{tabular}
	\label{tab:progression}
\end{table}

As can be seen, the perceived I/O time within the application falls to around one half of a second, and results in almost a 180X speedup, when using the optimal ADIOS2 configuration, virtually eliminating the I/O bottleneck that was observed at the onset when using the classic PnetCDF I/O method.

\subsection{ADIOS2 In-situ Processing}
As ADIOS2 is a data management library and not purely an file I/O library, ADIOS2's in-situ analysis capabilities were tested and applied to a weather forecasting pipeline. A 2 hour forecast of the CONUS 2.5km case was configured to output a history file every 30 simulation minutes, with the aim of the test to show the large decrease in total time-to-solution when using an in-situ pipeline with ADIOS2 versus the standard process-after-run method using the classic PnetCDF I/O option. 

The ADIOS2 SST engine was selected in the XML configuration file, which buffers and then transfers the requested data to a consumer (reader) over the network, instead of writing out a file to the file system. 

For data processing, a Python based analysis script was employed that plots a slice of the temperature field over the continental United States and plots an image similar to the one seen in Fig. \ref{fig:pipeline}. To facilitate a time-to-solution comparison, two versions of the data processing script were created, one that reads the data using the netcdf4-python API and one that uses the ADIOS2 high-level Python API. Of note, is that the ADIOS2 based script does not need to be altered to support in-situ processing, as the support is built-in when using the stepping mode with the Pythonic \emph{for fstep in adios2\_fh} directive.

Fig. \ref{fig:runtime} shows a run time progression of the ADIOS2 based end-to-end pipeline, compared to the traditional PnetCDF sequential pipeline, with the I/O, Compute and Initialization times extracted and analyzed from the WRF output files, and the post-processing time recorded manually for the PnetCDF case. The ADIOS2 based pipeline results show an almost constant block of compute as the perceived write time by the application is almost negligible, due to the internal buffering of the SST engine. The SST engine sends the data to the consumer in the background while the computation continues. Meanwhile, with the PnetCDF pipeline, the computation is interrupted for long periods as the file I/O takes place. Once the computation has finished, the PnetCDF post-processing script is run, which further extends the time to solution.

In total, the in-situ ADIOS2 approach using the SST engine is able to almost halve the time-to-solution compared to the legacy PnetCDF based approach, displaying significant value for time-sensitive operational weather forecast applications.

\begin{figure*}[htbp]
	\centerline{\includegraphics[width=500pt,keepaspectratio]{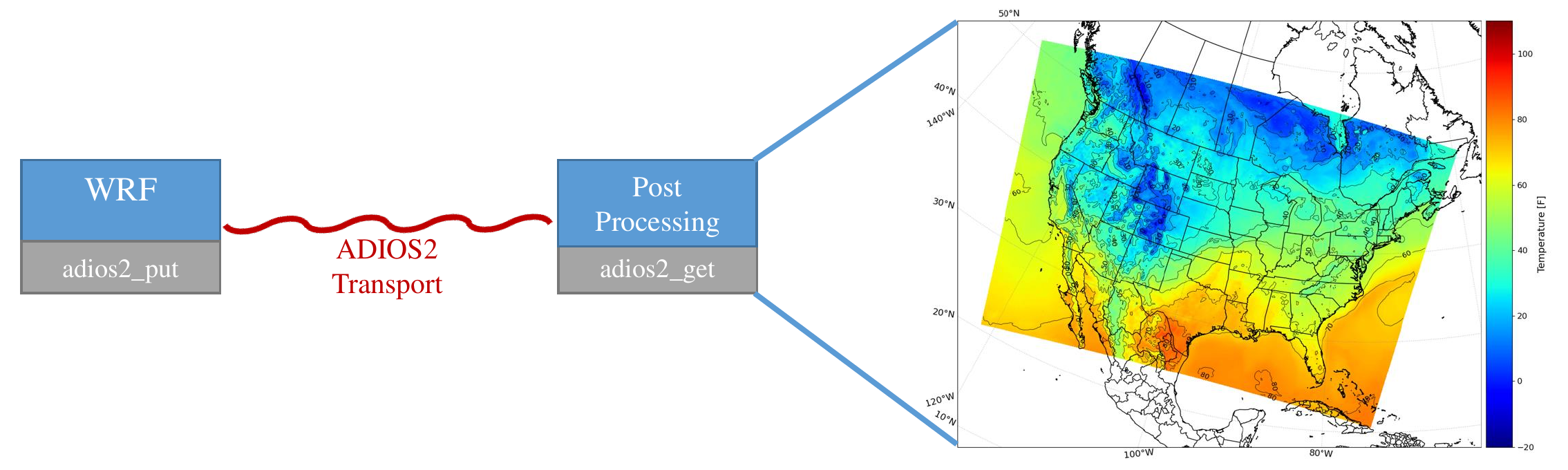}}
	\caption{Schematic of a WRF weather forecasting pipeline using ADIOS2 in-situ processing. Data is streamed from WRF to the consumer over the network, while bypassing the file system entirely.}
	\label{fig:pipeline}
\end{figure*}

\begin{figure*}[htbp]
	\centerline{\includegraphics[width=500pt,keepaspectratio]{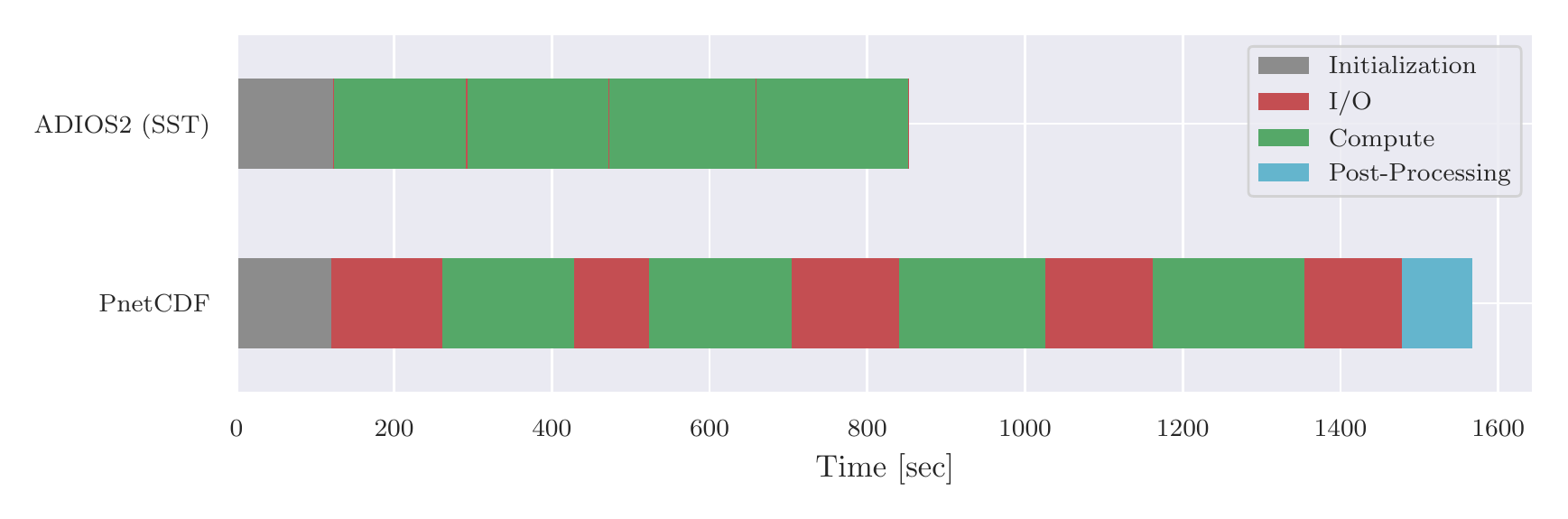}}
	\caption{Run time comparison of a WRF run with data processing. The ADIOS2 configuration processes the output data in-situ, using data streamed from WRF, while the PnetCDF configuration uses the traditional process-after-job-completion approach. }
	\label{fig:runtime}
\end{figure*}

Although not investigated here, the ADIOS2 data streaming engines open the door for new code-coupling possibilities for WRF, without the need to use the file system as a transfer mechanism, or develop application specific interfaces. 

\section{Conclusions}
\label{section:conclusions}
This work has presented the implementation and dramatic I/O performance improvements of adding the ADIOS2 data management library to WRF. The new library not only speeds up normal writes to PFS, but also adds a multitude of new features, such as node-local burst buffer write support, high-performance compression, possible two-way data coupling, and in-situ analysis. Tests results in this work indicate between one and two orders of magnitude improvement in perceived write time at scale when using the new ADIOS2 I/O backend. Additionally, a sample weather forecasting pipeline using the ADIOS2 SST engine was shown to decrease the total time to solution by half compared to the legacy PnetCDF method at scale. 

This work outlines the new capabilities and performance gains of the ADIOS2 library within WRF, tested on the benchmark CONUS 2.5km case. Future work will apply these new data streaming capabilities to large and research relevant WRF simulation cases that currently are constrained by poor I/O performance and slow, sequential analysis pipelines. Additionally, the effect of using lossy compression techniques for Numerical Weather Prediction should be investigated. The additional effective I/O throughput that can be achieved by lossy compression, versus the loss in numerical accuracy, needs to be carefully studied.


\balance
\bibliographystyle{IEEEtran}
\bibliography{IEEEabrv,./arxiv_2022_bib}
\end{document}